\documentclass[nofootinbib,aps,11pt]{revtex4}
\usepackage{graphicx}
\usepackage{cancel}
\usepackage{amssymb}
\usepackage{textcomp}
\usepackage{amsmath}
\usepackage{bm}
\usepackage{times}
\usepackage{epsfig}
\usepackage{color}
\usepackage{graphics}
\usepackage{hyperref}

\DeclareMathOperator{\tr}{tr}

\def\tr{\text{Tr}\,}
\begin{document}

\newcount\hour \newcount\minute
\hour=\time \divide \hour by 60
\minute=\time
\count99=\hour \multiply \count99 by -60 \advance \minute by \count99
\newcommand{\mydate}{\ \today \ - \number\hour :00}

\title{\huge \textit{The Fate of R-Parity}}
\author{Pavel Fileviez P\'erez, Sogee Spinner}
\affiliation{Department of Physics, University of Wisconsin-Madison, WI 53706, USA}
\date{\today}

\begin{abstract}
The possible origin of the R-parity violating interactions in the minimal supersymmetric 
standard model and its connection to the radiative symmetry breaking mechanism (RSBM) 
is investigated. In the context of the simplest model where the implementation of the 
RSBM is possible, we find that in the majority of  the parameter space R-parity is 
spontaneously broken at the low-scale. These results hint at the possibility that 
R-parity violating processes will be observed at the Large Hadron Collider, 
if Supersymmetry is realized in nature.
\end{abstract}

\maketitle
\section{Introduction}
The minimal supersymmetric standard model (MSSM) is considered as one of the most appealing extensions 
of the standard model of strong and electroweak interactions. This theory has a variety of appealing 
characteristics including solutions to the hierarchy problem and a dark matter candidate.  
However, at the renormalizable level, the MSSM Lagrangian contains flagrant baryon and lepton 
number violating operators, the most infamous of which lead to rapid proton decay 
(See Ref.~\cite{Haber} for a review on supersymmetry (SUSY) and Ref.~\cite{Nath} 
for the study of  the proton decay issue in SUSY.).

The most common approach to this problem is the introduction of a discrete symmetry, 
$R$-parity, defined as $R=(-1)^{3(B-L)+2S}$, where $B$, $L$ and $S$ are baryon and lepton number, 
and spin, respectively (See Ref.~\cite{Barbier:2004ez} for a review on $R$-parity violation.).
The conservation of $R$-parity has the added bonus of insuring that the lightest SUSY particle (LSP) 
is stable and therefore a cold dark matter candidate. While $R$-parity is closely linked to $B-L$, they 
are not synonymous.  Specifically, $R$-parity allows for terms that break $B-L$ by an even amount. 
For general arguments on $R$-parity conservation see Refs.~\cite{mohapatra} and \cite{Martin}.

In order to understand the conservation or violation of $R$-Parity one has to 
consider theories where $B-L$ is part of the gauge symmetry.  In such cases
$R$-parity is an exact symmetry as long as the same is true for $B-L$.  Breaking $B-L$
by a field with even charge (the canonical $B-L$ model) guarantees
automatic $R$-parity conservation even below the symmetry scale, since only $B-L$
violation by an even amount is allowed.  An alternative is $B-L$ breaking through
the right-handed sneutrino, a field which must always be included due to anomaly cancellation.  Since the
right-handed sneutrino has a charge of one, its VEV results in spontaneous $R$-parity violation.  
Phenomenologically, this is a viable scenario that does not induce tree-level rapid proton decay 
and dark matter is still possible if the gravitino is the LSP.  

Recently, spontaneous $R$-parity violation has been studied in the case of minimal $B-L$ models~\cite{paper1,paper2,paper3,paper4,paper5}.  
However, the following question is still relevant: {\textit{Does the canonical $B-L$ model favors $R$-parity conservation or violation?}}. 
In this letter we study this question in the simplest local $U(1)_{B-L}$ extension of the MSSM 
assuming for simplicity MSUGRA boundary conditions for the soft terms.  We investigate the 
fate of $R$-parity using the radiative symmetry breaking mechanism and show that 
for the majority of the parameter space, $R$-parity is broken, namely it is the right-handed sneutrino 
that acquires a negative mass squared and therefore a vacuum expectation value (VEV).  This is a 
surprising result that at the very least questions the feasibility of conserving $R$-parity in such a framework.
These results are quite general and apply to any SUSY theory where $B-L$ is part 
of the gauge symmetry.

\section{Theoretical Framework}
We investigate the possible connection between 
RSBM and the fate of $R$-parity in the simplest $B-L$ model,
based on the gauge group:
\begin{equation*}
SU(3) \bigotimes SU(2)_L \bigotimes U(1)_Y \bigotimes U(1)_{B-L}
\end{equation*}
with particle content listed in Table I.
\begin{table}[htdp]
\begin{center}
\begin{tabular}{|c|c|c|c|}
\hline
Field							& $SU(2)_L$	& $U(1)_Y$ & $U(1)_{B-L}$
\\
\hline
$\hat{Q} = \left(\hat{u}, \hat{d}\right)$	& 2			& 1/6          & 1/3
\\
$\hat{u}^c$						& 1			& -2/3         & -1/3
\\
$\hat{d}^c$					   & 1	  	 & 1/3         & -1/3
\\
$\hat{L} = \left(\hat{\nu}, \hat{e}\right)$ & 2		 & -1/2         & -1
\\
$\hat{e}^c$					   & 1	  	 & 1             & 1
\\
$\hat{\nu}^c$					   & 1	  	 & 0             & 1
\\
$\hat{H_u} = \left(\hat{H_u^+}, \hat{H_u^0}\right)$	& 2	& 1/2          & 0
\\
$\hat{H_d} = \left(\hat{H_d^0}, \hat{H_d^-}\right)$	& 2	& -1/2          & 0
\\
$\hat{X}$						   & 1	  	 & 0             & -2
\\
$\hat{\bar X}$					   & 1	  	 & 0             & 2
\\
\hline
\end{tabular}
\end{center}
\label{Charges}
\caption{$SU(2)_L \bigotimes U(1)_Y \bigotimes U(1)_{B-L}$ charges for the particle content.}
\end{table}

The most general superpotential is given by
\begin{align}
	{\cal W} = & \ {\cal W}_{MSSM} \ + \  {\cal W}_{B-L},
	\\
	{\cal W}_{MSSM} = & \ Y_u \ \hat{Q} \ \hat{H}_u \ \hat{u}^c 
		\ + \ Y_d \ \hat{Q} \ \hat{H}_d \  \hat{d}^c
		\ + \ Y_e \ \hat{L} \  \hat{H}_d \ \hat{e}^c
		\ + \ \mu \ \hat{H}_u \ \hat{H}_d,
	\\
	{\cal W}_{B-L} = & \ Y_\nu \ \hat{L} \ \hat{H}_u \ \hat{\nu}^c
		\ + \  f \ \hat{\nu}^c \ \hat{\nu}^c \ \hat{X}
	 - \ \mu_X \ \hat{X} \  {\hat {\bar X}},
\end{align}
and the corresponding soft SUSY breaking Lagrangian is
\begin{align}
	\label{L.Soft}
	-\mathcal{L_\text{Soft}}  \supset &
	\left(
		a_\nu \ \tilde L \ H_u \ \tilde \nu^c
		\ - \ a_X \ \tilde \nu^c \ \tilde \nu^c \ X
		\ - \ b_X \ X \ \bar X
	\  + \ \frac{1}{2} \ M_{BL} \tilde B' \tilde B' \ + \ h.c.
	\right)
	\notag
	\\
	& + m_X^2 |X|^2 + m_{\bar X}^2 |\bar X|^2 + m_{\tilde \nu^c}^2 |\tilde \nu^c|^2,
\end{align}
where we have suppressed flavor and group indices and $\tilde{B}'$ is the $B-L$ gaugino.

Spontaneous $B-L$ violation requires either the VEV of $X$, $\bar X$ or
$\tilde \nu^c$ to be nonzero, however the fate of $R$-parity lies solely in
the VEV of $\tilde \nu^c$: $\left<\tilde \nu^c\right> = 0$ corresponds to $R$-parity
conservation while $\left<\tilde \nu^c\right> \neq 0$ indicates spontaneous $R$-parity
violation.  Addressing the values of these VEVs requires the minimization conditions
which can be derived from the full potential where
$\left(\left<X \right>, \left<\bar X \right>, \left<\tilde \nu^c \right>\right)
= 1/\sqrt{2} \left(x, \bar x, n\right)$ \footnote{Technically, the left-handed sneutrino
has a VEV as well, but in order to generate the correct neutrino masses,
this VEV must be quite small compared to the others and so can safely be
ignored here~\cite{paper2}.}:
\begin{align}
	\left< V \right> = & \left< V_F\right> + \left< V_D\right> + \left< V_\text{Soft}\right>,
	\\ 
	\label{V.F}
	\left< V_F \right> = & \frac{1}{4} \ f^2 \ n^4
		\ + \ f^2 \ n^2 \ x^2 
		\ + \ \frac{1}{2} \ \mu_X^2 \left(x^2 + \bar x^2\right)
	 - \ \frac{1}{\sqrt 2} \ f \ \mu_X \ n^2 \ \bar x,
	\\
	\left< V_D \right> = & \frac{1}{32} \ g_{BL}^2 \left(2 \ \bar x^2 \ - \ 2 \ x^2 \ + \ n^2  \right)^2,
	\\ 
	\label{V.Soft}
	\left< V_{\text{Soft}} \right> = & -\frac{1}{\sqrt 2} \ a_X \ n^2 \ x
		\ - \ b_X x \ \bar x
		\ + \ \frac{1}{2} \ m_X^2 \ x^2 
		+ \ \frac{1}{2} \ m_{\bar X}^2 \ \bar x^2
		\ + \ \frac{1}{2} \ m_{\tilde \nu^c}^2 \ n^2.
\end{align}
Only two cases exist for spontaneous $B-L$ symmetry breaking:
Case \textit{i}) $n=0; \ x, \bar x \neq 0$ implying $R$-parity conservation or
Case \textit{ii}) $x, \bar x, n \neq 0$ implying spontaneous $R$-parity violation.
Note that a third case, $n \neq 0; \ x, \bar x = 0$ cannot exist due to the linear term for 
$x$ in Eq.~(\ref{V.Soft}) and for $\bar x$ in Eq.~(\ref{V.F}), which always induce a VEV for these fields.
\begin{itemize}
\item \underline{Case \textit{i})}: {\bf $R$-Parity Conservation} 

This is the traditional case studied in the literature.  The minimization 
conditions for $x$ and $\bar x$ are very similar in form 
to those of $v_u$ and $v_d$ in the MSSM:
\begin{align}
	\label{BL.min.cond}
	\frac{1}{2} M_{Z'}^2 & = -|\mu_X|^2
		\ + \ \frac
		{
			m_X^2 \ \tan^2 z \ - \ m_{\bar X}^2
		}
		{
			1 \ - \ \tan^2 z
		},
\end{align}
where $\tan z \equiv x/\bar x$ and
$M_{Z'}^2 \equiv g_{BL}^2 \left( x^2 + \bar x^2 \right)$, which
is the mass for the $Z'$ boson associated with broken $B-L$.

To attain a better understanding of the situation, let us examine Eq.
(\ref{BL.min.cond}) in the limit $x \gg \bar x$, with $m_X^2 < 0$ and
$m_{\bar X}^2 >0$, so that it reduces to
\begin{equation}
	\label{X.min.cond}
	\frac{1}{2} M_{Z'}^2 = -|\mu_X|^2 \ - \ m_X^2.
\end{equation}
Since the left-hand side is positive definite, \textit{the relationship $-m^2_X >  |\mu_X|^2$
must be obeyed for spontaneous $B-L$ violation}: a tachyonic $m^2_X$ is not enough.
This relationship between $\mu_X$ and $m_X$ is similar to the relationship in the MSSM
between $\mu$ and $m_{H_u}$ a relationship typically referred to as the $\mu$ problem,
\textit{i.e.} why is $\mu$ of the order of the SUSY mass scale.  Then in case \textit{i}, in
addition to the MSSM $\mu$ problem, we have introduced a new $\mu$ problem for
$\mu_X$.

As can be seen from Eq.~(\ref{X.min.cond}), $x$ is of order the SUSY mass scale
or about a TeV.  Replacing $X$ by its VEV in the term $f \nu^c \nu^c X$ in the superpotential
leads to the heavy Majorana mass term for the right-handed neutrinos and ultimately
to the Type I seesaw mechanism~\cite{TypeI} for neutrino masses:
\begin{equation}
	m_\nu = v_u^2 \ Y_\nu^T \ \left(f x\right)^{-1} \ Y_\nu.
\end{equation}
Since the mass of the right-handed neutrinos are of order TeV, realistic neutrino masses require, $Y_\nu \sim 10^{-6-7}$.  The rest of the spectrum is given in Appendix \ref{App.Spec}.

\item \underline{Case \textit{ii})}: {\bf $R$-Parity Violation} 

Evaluation of the minimization conditions in this case is illuminating in the limit $n \gg x, \ \bar x, a_X$ and $g_{BL}^2 \ll 1$, which will prove to be the case of interest in the numerical section:
\begin{align}
	n^2 = & \frac
	{
		\left(-m_{\tilde \nu^c}^2\right) \Lambda_{\bar X}^2
	}
	{
		f^2 \ m_{\bar X}^2
		\ + \ \frac{1}{8} \ g_{BL}^2
		\ \Lambda_{\bar X}^2
	},
	\\
	\label{min.cond.BL.ii}
	\bar x = & \frac
	{
		\left(-m_{\tilde \nu^c}^2\right) \ f \ \mu_X
	}
	{
		\sqrt{2}
		\left(
			f^2 \ m_{\bar X}^2
			\ + \ \frac{1}{8} \ g_{BL}^2 \ 
			\Lambda_{\bar X}^2
		\right)
	},
	\\
	x = &
		\frac
		{
			\left(-m_{\tilde \nu^c}^2\right)
			\left[
				a_X \Lambda_{\bar X}^2
				+ f \ b_X \ \mu_X
			\right]
		}
		{
			\left(2 \ f^2 - \frac{1}{4} g_{BL}^2\right) \left(-m_{\tilde \nu^c}^2\right) \Lambda_{\bar X}^2
			\ + \
			f^2 \ m_{\bar X}^2 \Lambda_{X}^2
			\ + \
			\frac{1}{8} g_{BL}^2 \Lambda_{\bar X}^2 \Lambda_{X}^2
		},
\end{align}
where $\Lambda_X^2 \equiv \mu_X^2 \ + \ m_X^2$ and
$\Lambda_{\bar X}^2 \equiv \mu_X^2 \ + \ m_{\bar X}^2$.

These equations indicate several things:
\textit{spontaneous $B-L$ symmetry breaking in the $R$-parity violating case
only requires $m_{\tilde \nu^c}^2 < 0$} and does not introduce a new $\mu$ problem 
so that $\mu_X$ can be larger than the TeV scale;
that $x$ and $\bar x$ are triggered by linear terms since they go as these linear
terms suppressed by the effective mass squared;
and all VEVs increase with $\mu_X$ up to a point after which $n$ asymptotes while $x$ and
$\bar x$ decrease as $1/\mu_x$.  The $\mu \to \infty$ serves as a decoupling limit since
$x, \ \bar x \to 0$ and $n^2 \to -8 m_{\tilde \nu^c}/g_{BL}^2$
as in the  minimal model~\cite{paper2}.
Neutrino masses in this case will have a more complicated form that will
depend both on the type I seesaw contribution and an $R$-parity contribution
although the bounds on $Y_\nu$ are similar to Case \textit{i}).  The $Z'$ mass
in this case is
\begin{equation}
	M_{Z'}^2 = \frac{1}{4} \left(n^2 \ + \ 4 \  x^2 \ + \ 4 \ \bar x^2\right).
\end{equation}
and the rest of the spectrum is given in Appendix \ref{App.Spec}.
\end{itemize}
The important question now becomes: are either of these cases possible
from the perspective of RSBM?  Specifically,
will running from some SUSY breaking boundary conditions drive either $X$ or
$\tilde \nu^c$ tachyonic, or neither.  To answer this we must turn to a specific SUSY breaking
scheme with some predictive power. One of the simplest way to transmit SUSY breaking 
is through gravity~~\cite{Chamseddine:1982jx} and here we will adopt the MSUGRA  
Ansatz with the following boundary conditions at the GUT scale:
\begin{align}
	m_X^2 &= m_{\bar X}^2 = m_{\tilde \nu_i^c}^2  = ... = m_0^2
	\\
	A_X &= f \ A_0; \ A_\nu = Y_\nu \ A_0; \ ...
	\\
	M_{BL} & =...= M_{1/2},
\end{align}
where $...$ indicates MSSM parameters.  

Finally, we present the renormalization group of equations (RGEs) necessary to evolve the boundary conditions given
by MSUGRA down to the SUSY scale, derived using~\cite{Martin:1993zk}.  The
RGEs will only be functions of the beyond the MSSM couplings since $Y_\nu$
is small enough to be neglected.  We assume that $g_{BL}$ unifies with the other
gauge couplings at the GUT scale of about $2 \times 10^{16}$ and for simplicity we 
use the SO(10) GUT renormalization factor, $\sqrt{3/8}$.  
In the one family approximation, the RGEs are given by\footnote{We would like to note that our results 
are in disagreement with the results in Ref. \cite{Masiero}.}
\begin{align}
	\label{beta.m.nu}
	16 \pi^2 \frac{d m_{\tilde \nu^c}^2}{d t} = & 
		\left[
			8 f^2 X_X
			 - \ 3 g_{BL}^2 \ M_{BL}^2
		\right],
	\\
	\label{beta.m.X}
	16 \pi^2 \frac{d m_X^2}{d t} = &
		\left[
			  4 f^2 X_X
		 - \ 12 g_{BL}^2 \ M_{BL}^2
		\right],
	\\
	\label{beta.m.Xbar}
	16 \pi^2 \frac{d m_{\bar X}^2}{d t} = & - {12 \ g_{BL}^2 \ M_{BL}^2},
\end{align}
where $t = ln \ \mu$, and $X_X \equiv m_X^2 + 2 m_{\tilde \nu^c}^2 + 4 a_X^2$.
See Appendix~\ref{App.RGE} for the full set of RGEs including the contributions from
three families of right-handed neutrinos.

Radiative symmetry breaking requires one of the soft masses
in Eqs.~(\ref{beta.m.nu} - \ref{beta.m.Xbar}) to run negative.  Experience from
radiative electroweak symmetry breaking in the MSSM~\cite{Ibanez:1982fr}, 
indicates that Yukawa terms in the beta functions tend to drive the masses squared
negative while gaugino terms do the opposite.  Due to its smaller $B-L$ charge,
$\tilde \nu^c$ has the smallest gaugino factor while also having the largest Yukawa factor.
Since in MSUGRA, all of these fields have the same mass at the GUT scale,
it is clear that $m_{\tilde \nu^c}^2$ will evolve to the smallest value in the simple one family
approximation.  When including all three values families, $m_X^2$ gets an enhancement from
trace of $f$, Eq.~(\ref{beta.m.X.full}), which could lead to it being tachyonic and therefore to $R$-parity conservation.  
The question of whether RSBM is possible
as well as the fate of $R$-parity throughout the parameter space
will be addressed numerically in the next section.

\section{R-Parity: Conservation or Violation ?}
In addition to addressing the feasibility of RSBM
in general and the fate of $R$-parity specifically,
it would also be prudent to identify the part
of parameter space that leads to a realistic spectrum.  One
strong experimental constrain is the bound on the $Z'$ mass:
$M_{Z'}/g_{BL} > 5$ TeV~\cite{Carena:2004xs}, indicating 
the need for a large mass scale, independent of the fate
of $R$-parity, and translates into a large value for
$m_0$ at the GUT scale.

Common lore dictates that a large mass scale at the GUT
scale also leads to large fine-tuning in the MSSM Higgs
sector.  However, large values of $m_0$ (TeV) and small
values of $M_{1/2}$ (few hundred GeV) (the so called
focus point region of MSUGRA~\cite{Feng:1999mn}), provides a remarkable
opportunity.  In this regime, the $H_u$ soft mass
runs slowly to small values that do not require a large amount of
fine-tuning while the larger symmetry factors for the Yukawa terms
in the $m_X^2$ and $m_{\tilde \nu^c}^2$ RGEs, Eq.~(\ref{beta.m.X.full},
\ref{beta.m.nu.full}), run these masses tachyonic faster and
naturally lead to a slight hierarchy between the electroweak
scale and the $B-L$ scale, as suggested by the hierarchy between
the $Z$ mass and the bounds on the $Z'$ mass.  This is
independent of the status of $R$-parity.  Aside from the stops,
the remaining soft scalar masses
do not run much and the approximations
made in the previous section for case \textit{ii} are valid.

The remainder of the parameters will be chosen as follows: $\tan \beta$
and $f$ values will be inputted at the SUSY scale.  Using these values,
the Yukawa couplings are evolved up to the GUT scale where $g_{BL}$
is assumed to unify with the other gauge couplings.  The MSUGRA parameters
are chosen in the focus point regime with $A_0 = 0$ (we find that $A_0$ has
very little effect on the results).  The SUSY breaking parameters are evolved down to the SUSY scale
where the EWSB minimization conditions are used to solve for $\mu$ and $B$.
It is also assumed that $B_X = B$ at the GUT scale, where $b_X = B_X \mu_X$.  Specifying
$\mu_X$ then determines the spectrum.

\begin{figure}[h!]
\begin{center}
\includegraphics[width=9cm]{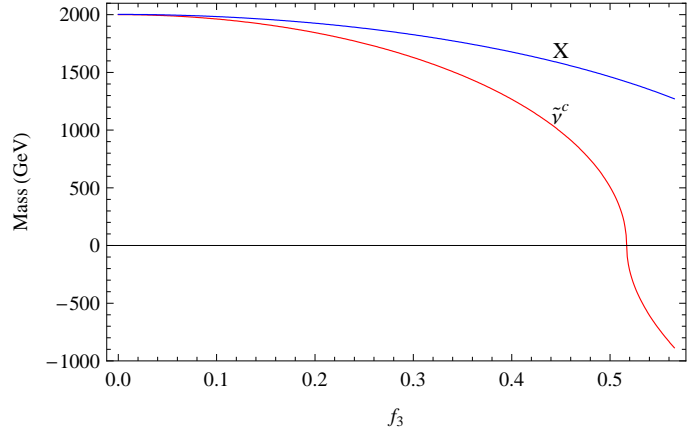}
\caption{Soft masses in the form $\text{sign}(m_\phi^2)|m_\phi|$ for
	$X$ (blue) and $\tilde \nu^c$ (red) versus $f_3$, for $m_0 = 2000$ GeV,
	$M_{1/2} = 200$ GeV, $A_0 = 0$ and negligible $f_1$ and $f_2$.  RSBM is possible for
	$f_3 \gtrsim 0.51$ and spontaneous $R$-parity violation.}
\label{Mass.v.f3}
\end{center}
\end{figure}

The feasibility of RSBM, as well as the fate of $R$-parity,
rely heavily on $f = \text{diag}\left(f_1, f_2, f_3\right)$.  
Calculating the soft masses of $X$ and $\tilde \nu^c$ with increasing $f_3$
yields Fig.~\ref{Mass.v.f3}, for $m_0 = 2000$ GeV,
$M_{1/2} = 200$ GeV, $A_0 = 0$ and negligible $f_1$ and $f_2$.  As expected, 
in the $f_1, f_2 \ll f_3$ limit,
only the $\tilde \nu^c$ mass becomes
tachyonic, so while RSBM
can be successful, it leads to spontaneous $R$-parity breaking.
Note that $f_3$ exhibits fixed-point like behavior (as
discussed in a similar scenario in \cite{Mohapatra:2008gz}).
This means that its range
allowing for RSBM, corresponds to a larger range of values
at the GUT scale.
\\

\begin{figure}[!h]
\begin{center}
\includegraphics[width=9.0cm]{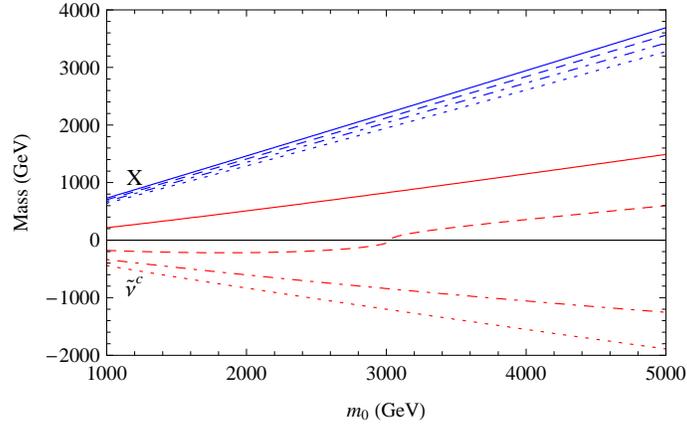}
\caption
{Soft masses in the form $\text{sign}(m_\phi^2)|m_\phi|$ for $X$ (blue) and $\tilde \nu^c$ (red)
	versus $m_0$
	for $f_3 = 0.5 \text{ (solid) }, 0.52 \text{ (dashed) }, 0.54 \text{ (dot-dashed) }, 0.56 \text{ (dotted)}$
	and all other parameters the same as in Fig.~\ref{Mass.v.f3}.}
\label{mnuc.m0.f}
\end{center}
\end{figure}

In Fig.~\ref{mnuc.m0.f}, are the $X$ and $\tilde \nu^c$ soft masses
for different values $f_3$ versus $m_0$ with all other parameters the
same as in Fig.~\ref{Mass.v.f3}.  It indicates that 
the $m_0$ parameter also plays an important role determining the overall
size of the tachyonic mass, and therefore the $Z'$ mass, 
and can even derail RSBM for lower values of $f_3$.

\begin{figure}[!h]
\begin{center}
\includegraphics[width=8.0cm]{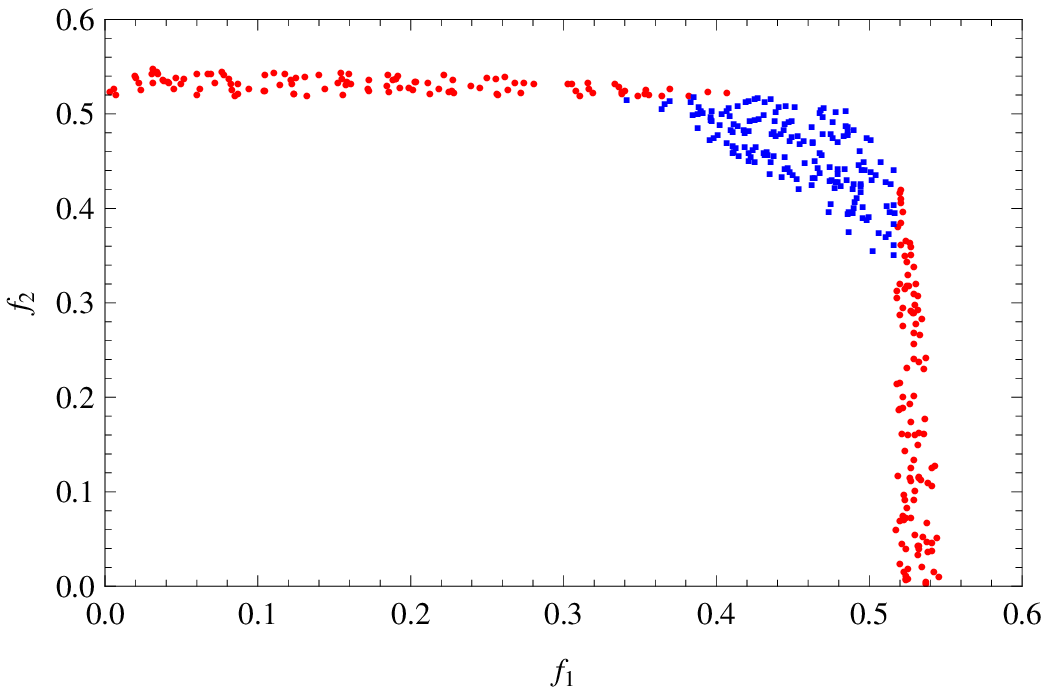}
\put(-8,45){(a)}
\includegraphics[width=8.0cm]{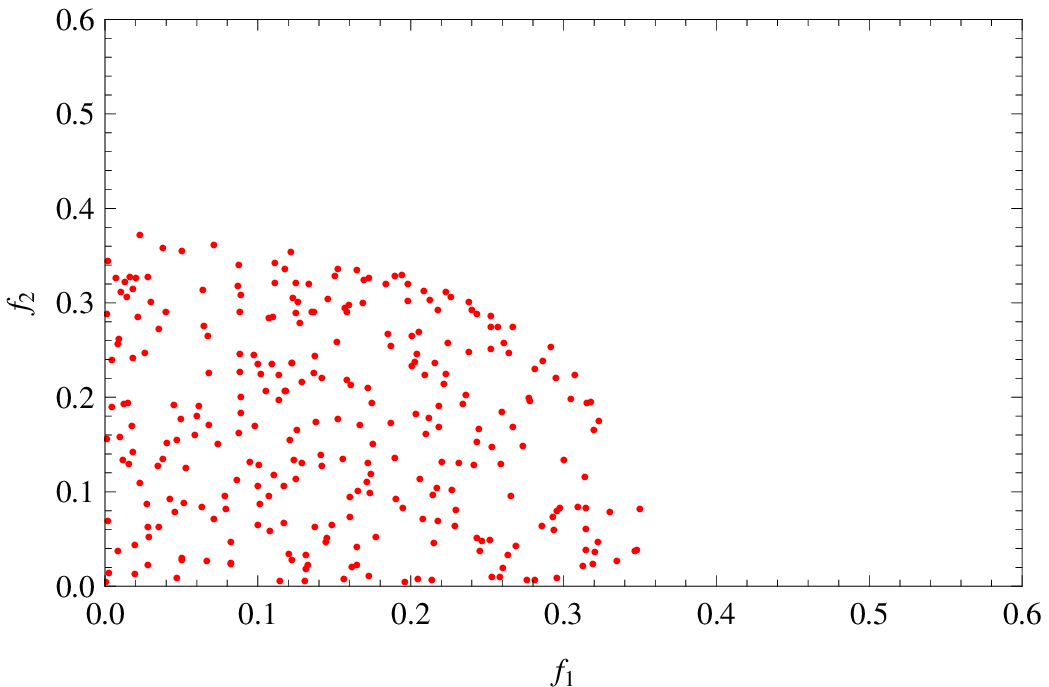}
\put(-8,45){(b)}
\caption
{The state of the $B-L$ breaking vacuum in the $f_2$--$f_1$ plane with $m_0 = 2000$ GeV,
	$M_{1/2} = 200$ GeV and $A_0 = 0$ for $f_3 = 0.4$ (a) and $f_3 = 0.55$ (b).  Blue dots indicate
	$R$-parity conservation while red dots $R$-parity violation.  In (a), the empty space
	below the curve indicates no RSBM, while in both graphs, in the space
	above the curves, the $f$'s are no longer perturbative at the GUT scale.  In (a),
	there is about an even number of $R$-parity conserving and violating vacua but increasing
	$f_3$ tips the favor towards $R$-parity violation and eventually only allows for $R$-parity violation
	as in (b).}
\label{RPC.V.RPV}
\end{center}
\end{figure}

For $f_1 \sim f_2 \sim f_3$, the Yukawa term in the RGE for $m_X^2$
is effectively enhanced by a factor of three, see Eq.~(\ref{beta.m.X.full})
as compared to Eq.~(\ref{beta.m.X}), which can lead to an $R$-parity conserving
minima since no such factor appears for 
$m_{\tilde \nu^c}^2$.  We show these effects in Fig.~\ref{RPC.V.RPV}, where red dots
indicate spontaneous $R$-parity violation and blue dots show the region
of $R$-parity conservation in the $f_2$--$f_1$ plane for $f_3 = 0.4$ (a)
and $f_3 = 0.55$ (b) and $m_0 = 2000$ GeV, $M_{1/2} = 200$ GeV and $A_0 = 0$.
In Fig.~\ref{RPC.V.RPV}(a) $f_1 \text{ or } f_2 \sim 0.52$ is needed for RSBM 
while only $f_1 \sim f_2 \gtrsim 0.4$ allows for $R$-parity conservation (there
is about a 50-50 split between $R$-parity conservation and violation in this graph).
If $f_1 \text{ or } f_2 > 0.52$, these couplings are no longer perturbative at the GUT scale.
As one increases the value of $f_3$, the $R$-parity conserving points disappear as
reflected in Fig.~\ref{RPC.V.RPV}(b), which does not allow for $R$-parity conservation.
In this case, $f_1 \text{ or } f_2  \gtrsim 0.4$ leads to non-perturbative values at the
GUT scale due to the larger value of $f_3$.
 
\begin{figure}[!ht]
\begin{center}
\includegraphics[width=9.0cm]{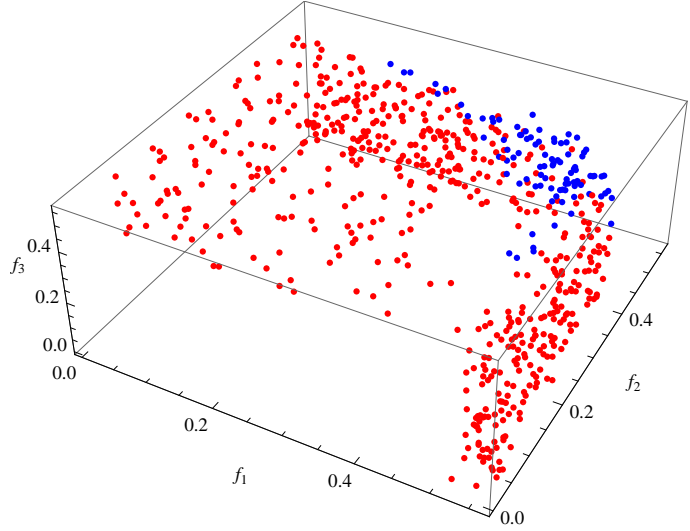}
\caption{The state of the $B-L$ breaking vacuum in the $f_1-f_2-f_3$ space
	with $m_0 = 2000$ GeV, $M_{1/2} = 200$ GeV and $A_0=0$.  Blue dots indicate
	$R$-parity conservation while red dots $R$-parity violation, the latter
	appears five times more often.  The key point is that only fairly degenerate
	values of $f$ (and therefore the right-handed neutrinos) allow for $R$-parity conservation.
	We have checked that all physical masses are positive in these cases.}
\label{3D1}
\end{center}
\end{figure}

\begin{figure}[!ht]
\begin{center}
\includegraphics[width=9.0cm]{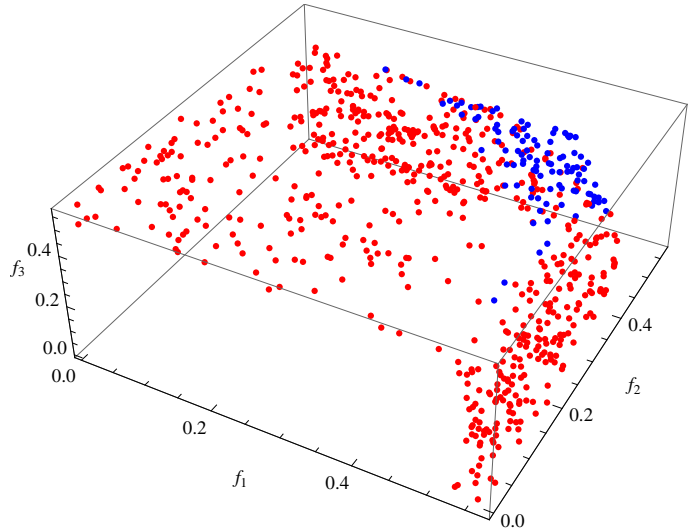}
\caption{The state of the $B-L$ breaking vacuum in the $f_1-f_2-f_3$ space
	with $m_0 = 5000$ GeV, $M_{1/2} = 500$ GeV and $A_0=0$.  Blue dots indicate
	$R$-parity conservation while red dots $R$-parity violation, the latter
	appears five times more often.  The key point is that only fairly degenerate
	values of $f$ (and therefore the right-handed neutrinos) allow for $R$-parity conservation.
	We have checked that all physical masses are positive in these cases.}
\label{3D2}
\end{center}
\end{figure}

The graphs in Fig.~\ref{RPC.V.RPV} are a bit misleading since they are just slices
of the three dimensional space $f_1-f_2-f_3$, which is displayed in Figs.~\ref{3D1} and \ref{3D2},
with the same legend as the former figure.  While
these latter figures are perhaps harder to read, one can see that the majority of the parameter
space which allows for RSBM is dominated by $R$-parity violation
(five times more prevalent) while only $f_1\sim f_2 \sim f_3$
allows for $R$-parity conservation.  Both of these regions sit on a thin shell where
$f_1 \text{ or } f_2 \text{ or } f_3 \sim 0.5$.  Below this shell, RSBM is
not realized.  This last figures summarize the findings of this letter quite well:
when RSBM is realized the $R$-parity breaking vacuum is 
more probable than the $R$-parity conserving one, especially when a hierarchy exists within
the $f$ matrix.  Only when this matrix is fairly degenerate (degenerate
right-handed neutrinos) does the running allow for $R$-parity conservation.  


\newpage
\section{Summary}
The possible origin of the $R$-parity violating interactions in the minimal extension of the standard model 
and its connection to the radiative symmetry breaking mechanism has been investigated in the simplest possible model.  
We have found that in the majority of the parameter space $R$-parity is spontaneously broken at the low-scale and the soft SUSY mass scale 
defines the $B-L$ and $R$-parity breaking scales. These results can be achieved in any extension 
of the MSSM where $B-L$ is part of the gauge symmetry. The main result of this letter hints at the possibility 
that $R$-parity violating processes will be observed at the Large Hadron Collider, if Supersymmetry 
is discovered. 
\subsection*{Acknowledgment}
{\small {We thank V. Barger, T. Han, X.Tata, M. J. Ramsey-Musolf, and N. Setzer 
for discussions. The work of P. F. P. was supported in part by the U.S. Department of Energy
contract No. DE-FG02-08ER41531 and in part by the Wisconsin Alumni
Research Foundation.  S.S. is supported in part by the U.S. Department 
of Energy under grant No. DE-FG02-95ER40896, and by the Wisconsin
Alumni Research Foundation.}}
\newpage
\appendix

\section{Renormalization Group Equations}
\label{App.RGE}
We present first the gamma functions, which
are useful for deriving the RGEs.  Here $i=1,2,3$:
\begin{align}
	\gamma_X = & \frac{1}{16 \pi^2}
		\left(
			2 \ \tr{f^2}  \ - \ 3 \ g_{BL}^2
		\right),
	\\
	\gamma_{\bar X} = & \frac{1}{16 \pi^2}
		\left(
			- 3 \ g_{BL}^2
		\right),
	\\
	\gamma_{\nu^c_i} = & \frac{1}{16 \pi^2}
		\left(
			4 \ f_i^2 \ - \ \frac{3}{4} \ g_{BL}^2
		\right),
\end{align}
where repeated indices are not summed and
$f = \text{diag}\left(f_1, \ f_2, \ f_3\right)$, since $f$
can always be diagonalized by rotating the right-handed
neutrino fields.  The same holds true here for $a_X$ due
to the MSUGRA Ansatz.

The RGEs are given by
\begin{align}
	16 \pi^2 \frac{d g_{BL}}{d t} = & {9 \ g_{BL}^3},
	\\
	\label{beta.f3}
	16 \pi^2 \frac{d f_i}{d t} = & {f_3}
		\left(
			8 \ f_i^2 \ + \ 2 \ \tr{f^2} \ - \ \frac{9}{2} \ g_{BL}^2
		\right),
	\\
	16 \pi^2 \frac{dM_{BL}}{dt} = &  {18 \ g_{BL}^2 M_{BL}},
	\\
	16 \pi^2 \frac{d a_{X_i}}{d t} = & 
		 \  {f_X}
		\left(
			16 \ f_i \ a_{X_i} \ + \ 4 \ \tr{\left(f \ a_X\right)} \ - \ 9 \ g_{BL}^2 \ M_{BL}
		\right)
		\\ &
	+ \  {a_{X_i}}
		\left(
			8 \ f_i^2 \ + \ 2 \ \tr{f^2} \ - \ \frac{9}{2} \ g_{BL}^2
		\right),
	\\
		16 \pi^2 \frac{d m_{\bar X}^2}{d t} = & - {12 \ g_{BL}^2 \ M_{BL}^2},
	\\
	\label{beta.m.X.full}
	16 \pi^2 \frac{d m_X^2}{d t} = & 
		\left[
			4 \ \tr {f^2} \ m_X^2
			  \ + 8 \ \tr{\left(f^2 m_{\tilde \nu^c}^2\right)}
			\ + \ 4 \ \tr{a_X^2}
		 - \ 12 g_{BL}^2 \ M_{BL}^2
		\right],
	\\
	\label{beta.m.nu.full}
	16 \pi^2 \frac{d m_{\tilde \nu^c_i}^2}{d t} = &
		\left[
			8 \ f_i^2  \left(m_X^2 \ + \ 2 \ m_{\tilde \nu^c_i}^2 \right)
			\ + \ 8 \ a_{X_i}^2
			 - \ 3 g_{BL}^2 \ M_{BL}^2
		\right].
\end{align}
\section{Spectrum}
\label{App.Spec}
In calculating the following spectrum we assume $\left<\tilde \nu^c_3, \ X, \ \bar X \right>
= \frac{1}{\sqrt 2} \left(n, \ x, \ \bar x \right)$ and all others zero.
Pseudoscalar mass matrix in the basis $\text{Im}\left(\tilde \nu^c_3, X, \bar X\right)$:
\begin{equation}
	\label{A.mass}
	\mathcal{M}_P = 
		\left( \begin{array}{ccc}
			2 \ \sqrt{2} \left(a_X \ x \ + \ f_3 \ \mu_X \ \bar x\right)
		&
			\sqrt{2} \ a_X \ n
		&
			-\sqrt{2} \ f_3 \ \mu_X \ n
		\\
			\sqrt{2} \ a_X \ n
		&
			\frac{a_X \ n^2 \ + \ \sqrt{2} \ b_X \ \bar x}{\sqrt{2} \ x}
		&
			b_X
		\\
			- \sqrt{2} \ f_3 \ n \ \mu_X
		&
			b_X
		&
			\frac{f_3 \ \mu_X \ n^2  \ + \ \sqrt{2} \ b_X \ x}{\sqrt{2} \ \bar x}
		\end{array}\right).
\end{equation}

Scalar mass matrix in the basis $\text{Re}\left(\tilde \nu^c_3, X, \bar X\right)$:
\begin{equation}
	\label{S.mass}
		\mathcal{M}_S = \left(\begin{array}{ccc}
			\left(2 \ f_3^2 \ + \ \frac{1}{4} \ g_{BL}^2\right) \ n^2
		&
			\left(4 \ f_3^2 \ - \ \frac{1}{2} \ g_{BL}^2\right) \ n \ x - \sqrt{2} \ a_X \ n
		&
			- \sqrt{2} \ f_3 \ \mu_x \ n \ + \ \frac{1}{2} g_{BL}^2 \ n \ \bar x
		\\
			\left(4 \ f_3^2 \ - \ \frac{1}{2} \ g_{BL}^2\right) \ n \ x - \sqrt{2} \ a_X \ n
		&
			\frac{a_3 \ n^2 \ + \ \sqrt{2} \ b_X \ \bar x}{\sqrt{2} \ x} \ + \ g_{BL}^2 \ x^2
		&
			-b_X \ - \ g_{BL}^2 \ x \ \bar x
		\\
			- \sqrt{2} \ f_3 \ \mu_x \ n \ + \ \frac{1}{2} g_{BL}^2 \ n \ \bar x
		&
			-b_X \ - \ g_{BL}^2 \ x \ \bar x
		&
			\frac{f_3 \ \mu_X \ n^2 \ + \ \sqrt{2} \ b_X \ x}{\sqrt{2} \ \bar x} \ + \ g_{BL}^2 \ \bar x^2
		\end{array}\right).
\end{equation}

Neutralino mass matrix in the basis $\left(B', \ \nu^c, \ \tilde X, \ \tilde{\bar X} \right)$:
\begin{equation}
	\mathcal{M}_{\chi^0} = \left(
	\begin{array}{cccc}
		M_{BL}
		&
		\frac{1}{2} \ g_{BL} \ n
		&
		-g_{BL} \ x
		&
		g_{BL} \ \bar x
	\\
		\frac{1}{2} \ g_{BL} \ n
		&
		\sqrt{2} \ f_3 \ x
		&
		\sqrt{2} \ f_3 \ n
		&
		0
	\\
		-g_{BL} \ x
		&
		\sqrt{2} \ f_3 \ n
		&
		0
		&
		-\mu_X
	\\
		g_{BL} \ \bar x
		&
		0
		&
		-\mu_X
		&
		0
	\end{array} \right)
\end{equation}
The sfermion mass, with matrices in the basis $\left(\tilde f_L, \ \tilde f_R\right)$
\begin{align}
	\mathcal{M}_{\tilde u}^2 =&
	\left(\begin{array}{cc}
		m_{\tilde Q}^2
		\ + \ m_{u}^2
		\ - \ \frac{1}{8} \left(g_2^2 \ - \ \frac{1}{3} \ g_1^2\right)\left(v_u^2 - v_d^2\right)
		\ + \ \frac{1}{3} D_{BL}
		&
		\frac{1}{\sqrt 2} \left(a_u \ v_u - Y_u \ \mu \ v_d\right)
	\\
		\frac{1}{\sqrt 2} \left(a_u \ v_u - Y_u \ \mu \ v_d\right)
		&
		m_{\tilde u^c}^2
		\ + \ m_{u}^2
		\ - \ \frac{1}{6} \ g_1^2 \left(v_u^2 \ - \ v_d^2\right)
		\ - \ \frac{1}{3} D_{BL}
	\end{array}\right),
	\\
	\mathcal{M}_{\tilde d}^2 =&
	\left(\begin{array}{cc}
		m_{\tilde Q}^2
		\ + \ m_{d}^2
		\ +\ \frac{1}{8} \left(g_2^2 \ + \ \frac{1}{3} \ g_1^2\right)\left(v_u^2 - v_d^2\right)
		\ + \ \frac{1}{3} D_{BL}
		&
		\frac{1}{\sqrt 2} \left(Y_d \ \mu \ v_u - a_d \ v_d\right)
	\\
		\frac{1}{\sqrt 2} \left(Y_d \ \mu \ v_u - a_d \ v_d\right)
		&
		m_{\tilde d^c}^2
		\ + \ m_{d}^2
		\ + \ \frac{1}{12} \ g_1^2 \left(v_u^2 \ - \ v_d^2\right)
		\ - \ \frac{1}{3} D_{BL}
	\end{array}\right),
	\\
	\mathcal{M}_{\tilde e}^2 =&
	\left(\begin{array}{cc}
		m_{\tilde L}^2
		\ + \ m_{e}^2
		\ +\ \frac{1}{8} \left(g_2^2 \ - \ g_1^2\right)\left(v_u^2 - v_d^2\right)
		\ - \ D_{BL}
		&
		\frac{1}{\sqrt 2} \left(Y_e \ \mu \ v_u - a_e \ v_d\right)
	\\
		\frac{1}{\sqrt 2} \left(Y_e \ \mu \ v_u - a_e \ v_d\right)
		&
		m_{\tilde e^c}^2
		\ + \ m_{e}^2
		\ + \ \frac{1}{4} \ g_1^2 \left(v_u^2 \ - \ v_d^2\right)
		\ + \ D_{BL}
	\end{array}\right),
\end{align}
\begin{align}
	m_{\tilde \nu_L}^2 =& \ m_{\tilde L}^2
		\ - \ \frac{1}{8} \left(g_2^2 \ + \ g_1^2\right)\left(v_u^2 \ - \ v_d^2 \right)
		\ - \ D_{BL},
	\\
	\label{m.NI}
	m_{\tilde N_{I_i}}^2 = & m_{\tilde \nu_i^c}^2
		\ + \ 2 f_i^2 \ x^2
		\ - \ f_i \ f_3 \ n^2
		\ + \ \sqrt 2 \ a_{X_i} \ x
		\ + \ \sqrt 2 \ f_i \ \mu_X \ \bar x
		\ + \ D_{BL},
	\\
	\label{m.NR}
	m_{\tilde N_{R_i}}^2 = & m_{\tilde \nu_i^c}^2
		\ + \ 2 f_i^2 \ x^2
		\ + \ f_i \ f_3 \ n^2
		\ - \ \sqrt 2 \ a_{X_i} \ x
		\ - \ \sqrt 2 \ f_i \ \mu_X \ \bar x
		\ + \ D_{BL}.
\end{align}
where $D_{BL} \equiv \frac{1}{8} \ g_{BL}^2 \left(2 \ \bar x^2 \ - \ 2\  x^2  \ + \ n^2\right)$,
and $m_u, \ m_d$ and $m_e$ are the respective fermion masses and $a_u, \ a_d$ and $a_e$
are the trilinear $a$-terms corresponding to the Yukawa couplings $Y_u, \ Y_d$ and $Y_e$.
The right-handed sneutrino eigenstates are the scalars $\tilde N_{R_i}$ and pseudoscalars
$\tilde N_{I_i}$ where $i$ runs only over the first two generations and repeated indices
are not summed.  The third generation mixes with the Higgses, Eqs.~(\ref{A.mass}, \ref{S.mass}).
The above masses are for $R$-parity violation, case \textit{ii} from the text.  For the 
$R$-parity conserving case, case \textit{i}, take the limit $n \to 0$ and the $B-L$ Higgs masses are given by the lower
two-by-two block matrices of Eqs.~(\ref{A.mass}, \ref{S.mass}) and $i$ in Eqs.~(\ref{m.NI}, \ref{m.NR})
runs over all three generations.



\begin{thebibliography}{99}

\bibitem{Haber}
  M.~F.~Sohnius,
  ``Introducing Supersymmetry,''
  Phys.\ Rept.\  {\bf 128} (1985) 39;
  S.~P.~Martin,
  ``A Supersymmetry Primer,''
  arXiv:hep-ph/9709356.  
  
\bibitem{Nath}
  P.~Nath and P.~Fileviez Perez,
  ``Proton stability in grand unified theories, in strings, and in branes,''
  Phys.\ Rept.\  {\bf 441} (2007) 191
  [arXiv:hep-ph/0601023].

\bibitem{Barbier:2004ez}
  R.~Barbier {\it et al.},
  ``R-parity violating supersymmetry,''
  Phys.\ Rept.\  {\bf 420}, 1 (2005)
  [arXiv:hep-ph/0406039].

\bibitem{mohapatra}
R.~N.~Mohapatra,
  ``New contributions to neutrinoless double-beta decay in supersymmetric
  theories,''
  Phys.\ Rev.\  D {\bf 34}, 3457 (1986).

\bibitem{Martin}
  A.~Font, L.~E.~Ibanez and F.~Quevedo,
  ``Does Proton Stability Imply the Existence of an Extra Z0?,''
  Phys.\ Lett.\  B {\bf 228} (1989) 79;
 S.~P.~Martin,
 ``Some simple criteria for gauged R-parity,''
  Phys.\ Rev.\  D {\bf 46} (1992) 2769
  [arXiv:hep-ph/9207218];
  ``Implications of supersymmetric models with natural R-parity conservation,''
  Phys.\ Rev.\  D {\bf 54} (1996) 2340.

\bibitem{paper1}
  P.~Fileviez Perez and S.~Spinner,
  ``Spontaneous R-Parity Breaking and Left-Right Symmetry,''
  Phys.\ Lett.\  B {\bf 673}, 251 (2009).
  [arXiv:0811.3424 [hep-ph]].
 
\bibitem{paper2}  
  V.~Barger, P.~Fileviez Perez and S.~Spinner,
  ``Minimal gauged U(1)-{B-L} model with spontaneous R-parity violation,''
  Phys.\ Rev.\ Lett.\  {\bf 102}, 181802 (2009).
 [arXiv:0812.3661 [hep-ph]].

\bibitem{paper3}
  P.~Fileviez Perez and S.~Spinner,
  ``Spontaneous R-Parity Breaking in SUSY Models,''
  Phys.\ Rev.\  D {\bf 80}, 015004 (2009).
 [arXiv:0904.2213 [hep-ph]].

\bibitem{paper4}  
  L.~L.~Everett, P.~Fileviez Perez and S.~Spinner,
  ``The Right Side of Tev Scale Spontaneous R-Parity Violation,''
  Phys.\ Rev.\  D {\bf 80}, 055007 (2009).
 [arXiv:0906.4095 [hep-ph]].

\bibitem{paper5}  
  P.~Fileviez Perez and S.~Spinner,
  ``TeV Scale Spontaneous R-Parity Violation,''
  AIP Conf.\ Proc.\  {\bf 1200}, 529 (2010).
 [arXiv:0909.1841 [hep-ph]].

\bibitem{TypeI}
  P.~Minkowski,
  ``Mu $\to$ E Gamma At A Rate Of One Out Of 1-Billion Muon Decays?,''
  Phys.\ Lett.\ B {\bf 67} (1977) 421;
  T. Yanagida,
in {\it Proceedings of the Workshop on the Unified Theory
   and the Baryon Number in the Universe}, eds. O. Sawada et al.,
p.~95, KEK Report 79-18, Tsukuba (1979);
  M. Gell-Mann, P. Ramond and R. Slansky,
   in {\it Supergravity}, eds. P. van Nieuwenhuizen et al.,
   (North-Holland, 1979), p.~315;
  S.L. Glashow, in {\it Quarks and Leptons}, Carg\`ese, eds. M. L\'evy et al.,
(Plenum, 1980), p. 707;
  R.~N.~Mohapatra and G.~Senjanovi\'c,
  ``Neutrino Mass And Spontaneous Parity Nonconservation,''
  Phys.\ Rev.\ Lett.\  {\bf 44} (1980) 912.


\bibitem{Chamseddine:1982jx}
  A.~H.~Chamseddine, R.~L.~Arnowitt and P.~Nath,
  ``Locally Supersymmetric Grand Unification,''
  Phys.\ Rev.\ Lett.\  {\bf 49}, 970 (1982);
  R.~Barbieri, S.~Ferrara and C.~A.~Savoy,
  ``Gauge Models With Spontaneously Broken Local Supersymmetry,''
  Phys.\ Lett.\  B {\bf 119}, 343 (1982);
  L.~E.~Ibanez,
  ``Locally Supersymmetric SU(5) Grand Unification,''
  Phys.\ Lett.\  B {\bf 118}, 73 (1982);
  L.~J.~Hall, J.~D.~Lykken and S.~Weinberg,
  ``Supergravity As The Messenger Of Supersymmetry Breaking,''
  Phys.\ Rev.\  D {\bf 27}, 2359 (1983);
  N.~Ohta,
  ``Grand Unified Theories Based On Local Supersymmetry,''
  Prog.\ Theor.\ Phys.\  {\bf 70}, 542 (1983).


\bibitem{Martin:1993zk}
  S.~P.~Martin and M.~T.~Vaughn,
  ``Two Loop Renormalization Group Equations For Soft Supersymmetry Breaking
  Couplings,''
  Phys.\ Rev.\  D {\bf 50}, 2282 (1994)
  [Erratum-ibid.\  D {\bf 78}, 039903 (2008)].
 [arXiv:hep-ph/9311340].


\bibitem{Masiero}
  S.~Khalil and A.~Masiero,
  ``Radiative B-L symmetry breaking in supersymmetric models,''
  Phys.\ Lett.\  B {\bf 665}, 374 (2008).
  [arXiv:0710.3525 [hep-ph]].

\bibitem{Ibanez:1982fr}
L.~Alvarez-Gaume, J.~Polchinski and M.~B.~Wise,
  ``Minimal Low-Energy Supergravity,''
  Nucl.\ Phys.\  B {\bf 221}, 495 (1983);
  L.~E.~Ibanez and G.~G.~Ross,
  ``SU(2)-L X U(1) Symmetry Breaking As A Radiative Effect Of Supersymmetry
  Breaking In Guts,''
  Phys.\ Lett.\  B {\bf 110}, 215 (1982).
  
\bibitem{Carena:2004xs}
  M.~S.~Carena, A.~Daleo, B.~A.~Dobrescu and T.~M.~P.~Tait,
  ``$Z^\prime$ gauge bosons at the Tevatron,''
  Phys.\ Rev.\  D {\bf 70}, 093009 (2004)
  [arXiv:hep-ph/0408098].

\bibitem{Feng:1999mn}
  J.~L.~Feng, K.~T.~Matchev and T.~Moroi,
  ``Multi-TeV scalars are natural in minimal supergravity,''
  Phys.\ Rev.\ Lett.\  {\bf 84}, 2322 (2000)
  [arXiv:hep-ph/9908309];
  J.~L.~Feng, K.~T.~Matchev and T.~Moroi,
  ``Focus points and naturalness in supersymmetry,''
  Phys.\ Rev.\  D {\bf 61}, 075005 (2000).
  [arXiv:hep-ph/9909334].

\bibitem{Mohapatra:2008gz}
  R.~N.~Mohapatra, N.~Setzer and S.~Spinner,
  ``Seesaw Extended MSSM and Anomaly Mediation without Tachyonic Sleptons,''
  JHEP {\bf 0804}, 091 (2008).
[arXiv:0802.1208 [hep-ph]].

\end{thebibliography}
\end{document}